# Vibration Analysis and Mitigation in Semiconductor Motion Stages Using DMAIC Methodology: A Case Study

**Yin Li**
Thrust of Data Science and Analytics
The Hong Kong University of Science and Technology (Guangzhou), Guangzhou, China
Tel: (+86) 20 88331234, Email: yligt@connect.hkust-gz.edu.cn

**Hua Chen**
Sino-German College of Intelligent Manufacturing
Shenzhen Technology University, Shenzhen, China
Tel: (+86) 755 23256198, Email: chenhua@sztu.edu.cn

**Fugee Tsung †**
Department of Industrial Engineering and Decision Analytics
The Hong Kong University of Science and Technology, Hong Kong SAR, China
Thrust of Data Science and Analytics
The Hong Kong University of Science and Technology (Guangzhou), Guangzhou, China
Tel: (+86) 20 88331234, Email: season@ust.hk

**Abstract.** Motion stages are critical in semiconductor manufacturing equipment for processes like die bonding, wafer loading, and chip packaging, as their performance must meet the industry's stringent precision requirements. Vibration, a significant yet often overlooked adversary to precision motion stages, is challenging to identify and mitigate due to its subtle nature. This study, conducted at a motion stage manufacturer facing frequent vibration-related complaints, proposes a novel approach to resolving vibration issues. By leveraging the DMAIC methodology, it introduces VIBGUARD, an active vibration monitoring and mitigation solution, instead of solely focusing on traditional hardware vibration control. This comprehensive strategy enhances value and competitiveness, increasing UPH (units per hour) by 15.3% from 8,500 to 9,800 and reducing downtime by 68.2% from 2.2 to 0.7 occurrences per month. This case study and the DMAIC methodology offer valuable resources for quality control and problem analysis in the semiconductor industry.

## 1. INTRODUCTION

Motion stages are essential in semiconductor manufacturing, significantly influencing equipment efficiency and precision in both front-end and back-end processes. High-speed, high-precision stages, as highlighted by Li et al. (2024) and Yazan et al. (2021), enhance production throughput and ensure product quality. State-of-the-art systems achieve speeds of 1 m/s, accelerations of 20g, and positioning precisions of ±200 nm, as illustrated in Figure 1. However, motion generates vibrations that can severely impact semiconductor processes, sensitive to disturbances from motors, fans, pumps, and external sources like traffic or seismic activity. Mitigation strategies, including passive, active, and semi-active control systems, have been developed (Huang and Xu, 2023). A major manufacturer, X, faced client complaints about vibrations affecting production yield and causing downtime, particularly in die-bonding systems where vibrations delayed camera capture times and reduced productivity.

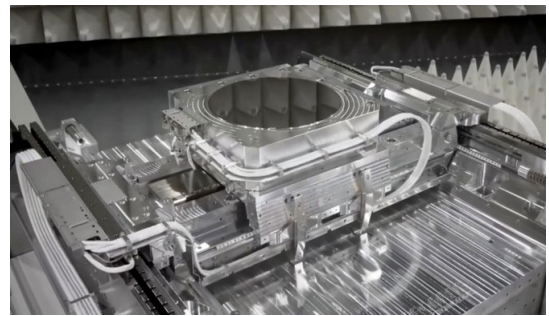

Figure 1: Motion stage in the semiconductor industry

Previous methods for vibration detection, control, and cancellation have focused on specific causes and solutions (Zheng et al., 2023) or predictive maintenance (King & Curran, 2019). However, motion stage manufacturer X faced a unique challenge due to its distance from the production environment, complicating the direct identification of vibration sources. To address this and meet end users' goals of increasing UPH by 12-15% and minimizing defects, a DMAIC (Define, Measure, Analyze, Improve, Control) methodology was employed to simulate field conditions and the production process. This approach, novel in this context, aimed to identify primary vibration causes, establish correlations, and implement effective control and monitoring measures. Successful implementation led to a significant increase in UPH and reduced downtime at the end-user level. Moreover, this methodology provides a proactive framework for quickly identifying and mitigating future unexpected vibration incidents.

## 2. LITERATURE REVIEW

### 2.1 DMAIC in Semiconductor Industry

The DMAIC methodology has proven effective for continuous process improvement in the semiconductor industry, driving significant enhancements in quality and efficiency. For instance, Tong et al. (2004) used DMAIC to improve printed circuit board (PCB) quality by addressing key production factors, resulting in better manufacturing processes and product quality. Similarly, Psi Technologies Inc. applied DMAIC to die separation processes, significantly reducing defects by identifying and correcting root causes (Curbano et al., 2020). Additionally, Gangidi (2019) utilized DMAIC to improve yield in semiconductor wafer fabrication by optimizing critical process factors, leading to notable yield improvements. Integrating statistical learning, machine learning, and AI with Six Sigma and DMAIC methodologies significantly enhances process optimization through data-driven insights (Tsung and Wang, 2023).

### 2.2 Vibration Analysis and Control

Vibration analysis and control in semiconductor machines and precision motion stages ensure high precision and stability in advanced manufacturing. Various methods, including passive isolation, active control systems, and input shaping, have been explored. Gupta et al. (2020) demonstrated friction isolators' effectiveness in mitigating vibrations. Zheng et al. (2020) developed a two-dimensional vibration stage with closed-loop control for micro milling. Aribowo et al. (2011) applied input shaping to reduce residual vibrations in wafer transfer robots, improving settling and movement times.

Current research lacks a comprehensive approach to utilizing DMAIC and Six Sigma for vibration analysis and control in the semiconductor industry. Specifically, there is a gap in providing integrated solutions for active vibration suppression, vibration source analysis, motion control, and monitoring, rather than relying solely on hardware improvements.

## 3. RESEARCH METHODS

Current vibration control methods in semiconductor equipment focus on hardware improvements and passive damping, which increase complexity and cost without addressing root causes. Passive damping is less effective in varying conditions and high-frequency ranges, and lacks real-time adaptability. Integrated monitoring and control are also missing, preventing timely adjustments. Therefore, a comprehensive approach combining active vibration suppression, real-time monitoring, and precise control is needed for optimal performance.

This research used the Six Sigma DMAIC methodology: Define, Measure, Analyze, Improve, and Control. In the Define stage, the problem and Critical to Quality (CTQ) factors were identified (Zaman et al., 2013). In the Measure stage, the operating environment of a motion stage in semiconductor equipment was simulated using Acceleration RMS as the vibration level indicator, with real-time measurement. In the Analyze stage, a fishbone diagram identified the three most impactful and controllable vibration causes. In the Improve stage, experiments using DOE and linear regression quantified these factors' impact on process capability (Jirasukprasert et al., 2014), and active vibration suppression was implemented. In the Control stage, a sensing and control scheme was proposed to monitor, analyze, and control vibrations with feedback mechanisms and alarms.

Active vibration suppression uses input shaping, a feedforward control technique that mitigates residual vibration and shortens system settling time by convolving the original control command with a shaper, as shown in Figure 2. Effective input shaping relies on accurate system modeling.

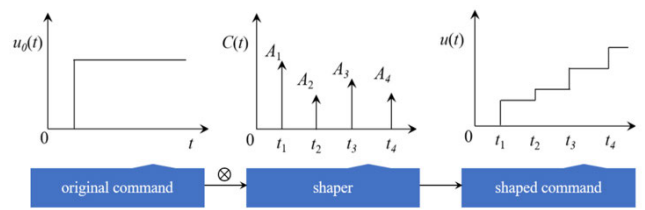

Figure 2: Shaping an original command with four impulses

To eliminate residual vibration, Equation (1) must equal zero, defining the Zero Vibration (ZV) shaper. Additional constraints in Equation (2) ensure a definite solution.

$$(\omega_n, \xi) = e^{-\xi \omega_n t} \sqrt{\left(\sum_{i=1}^{n} A_i e^{-\xi \omega_n t_i} \cos \omega_d t_i\right)^2 + \left(\sum_{i=1}^{n} A_i e^{-\xi \omega_n t_i} \sin \omega_d t_i\right)^2} \quad (1)$$

$$\begin{cases} \sum_{i=0}^{n} A_i = 1 \\ t_1 = 0 \end{cases} \quad (2)$$

Where $A_i$ and $t_i$ are amplitude and time of each impulse in shaper, and $\omega_n$ and $\xi$ are natural frequency and damping ratio respectively, and $\omega_d$ is damped frequency.

Several input shaping techniques, including ZVD, ZVDD, and EI, offer enhanced robustness compared to the ZV shaper. Based on a comparative analysis of their performance, the EI shaper is selected for active vibration suppression in this study.

## 4. CASE STUDY AND DISCUSSION

X Corporation, a leading precision motion stage manufacturer, supplies key components to the semiconductor equipment industry. These stages are integrated into machines used in semiconductor fabrication plants (FABs) for IC production. In this critical environment, even minor vibrations can cause costly defects and downtime.

The supply chain has two tiers: X Corporation supplies stages to equipment manufacturers, who deliver complete machines to FABs. When FABs face defects or downtime, it's challenging for X Corporation to quickly identify the root cause. The goal is to minimize defects and maximize units per hour (UPH), but troubleshooting is complex.

To aid problem analysis, a machine demonstration environment simulates real-world production conditions. This setup includes:

- Soft-motion controller: Controls the IO and motion of stages and utilizes EtherCAT fieldbus for communication with IO modules and servo drives.
- Two EtherCAT servo drives: Control two linear motors integrated into the motion stage.
- EtherCAT sensor module: Interfaces with multiple accelerometers placed on the stage endpoint and machine body to measure vibration.
- External Vibrator: Simulates external vibrations.

Figures 3 and 4 show the physical layout and system diagram of the demonstration setup.

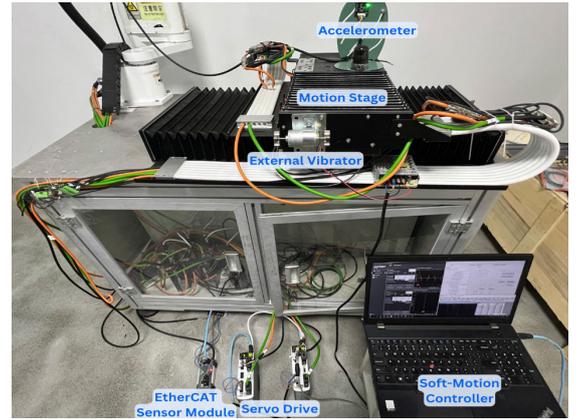

Figure 3: Physical layout of the demonstration

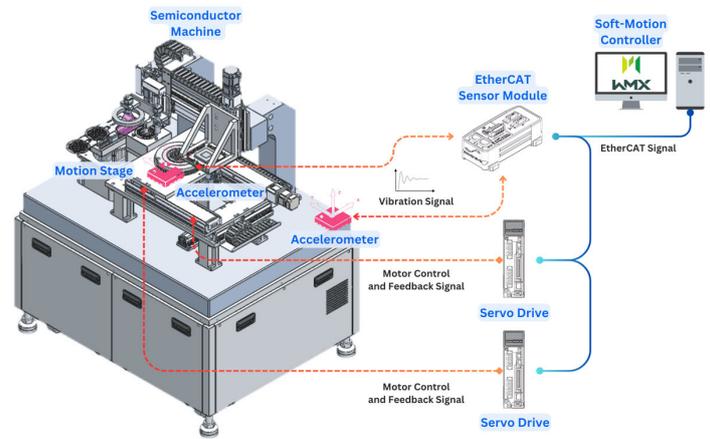

Figure 4: System diagram of the demonstration

### 4.1 Define Stage

To address the vibration challenge, the define stage of DMAIC involves clearly identifying the problem and setting objectives for the analysis.

Although X Corporation is not a direct supplier to end users (FABs), they must identify the Critical-to-Quality (CTQ) factors that impact the process and support the FAB's goals of increasing units per hour (UPH) by 12-15% and minimizing defects. Using a two-layered SIPOC (Suppliers-Inputs-Process-Outputs-Customers) diagram (Figure 5), X Corporation can analyze its role within the supply chain.

SIPOC reveals a clear chain: FABs request increased UPH and defect reduction from machine makers, who request specific motion specifications and accuracy from stage manufacturers like X Corporation. This process usually operates smoothly until FABs identify excessive vibration as the root cause, with the motion stage (supporting the wafer or chip) becoming the prime suspect.

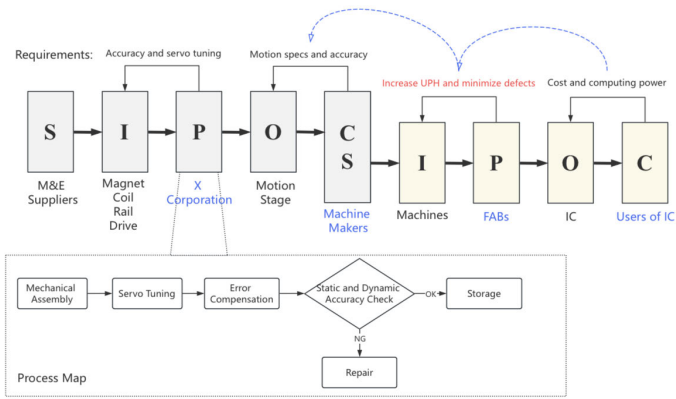

Figure 5: Two-layered SIPOC diagram

X Corporation faces a significant issue: its current process map lacks a vibration evaluation. The specifications only clearly define positioning accuracy, repeatability, payload, resolution, etc. Our demonstration setup and DMAIC methodology will help us solve this puzzle.

### 4.2 Measure Stage

In the Measure stage of DMAIC, data collection establishes a performance baseline and quantifies the problem. The UPH in dedicated processing within FABs fluctuates around 8500. Attempts to increase speed and acceleration to boost UPH cause endpoint displacement vibrations, leading to extended wait times for stable imaging or positioning within a <5μm window.

We used a Soft-motion controller to send positioning commands with various combinations of position, velocity, and acceleration. We collected command, feedback data from the controller, and vibration data from an accelerometer at the endpoint. The vibrations' peak, crest factor, and RMS were measured, with RMS being the primary vibration energy indicator, provided as Equation (3).

Active vibration suppression using EI input shaping was compared to the baseline. Results show significant endpoint vibrations depending on acceleration settings and distance from the stage's geometric center. Input shaping effectively mitigates these vibrations, as shown in Figure 6. Command and feedback data from the controller showed no vibration, indicating adequate servo tuning and no motor-side vibration, necessitating the accelerometer's placement at the endpoint.

$$RMS = \sqrt{\left(\frac{1}{n}\right)\sum_{i=1}^{n}(a_i)^2} \qquad (3)$$

Where $a_i$ represents the acceleration value measured at the i-th time point.

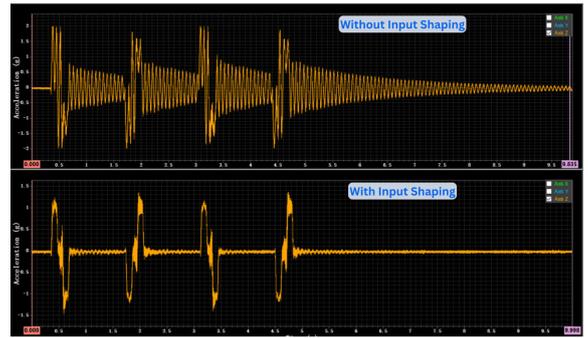

Figure 6: Vibration without and with input shaping

### 4.3 Analyze Stage

In the Analyze stage of DMAIC, identifying root causes of vibration-related defects is crucial. Using a Cause and Effect Diagram shown in Figure 7, several factors were identified: inadequate vibration suppression, mechanical issues (e.g., insufficient stiffness, poor design, lack of maintenance, component failures), electronic noise, external vibrations, and electrical control problems (e.g., improper controller commands, jerk, feedforward issues, poor PID/servo tuning).

After discussions with X corporation stakeholders, mechanical issues, controller commands, and electronic noise interference were ruled out due to prior troubleshooting. The remaining suspects are external vibrations, input shaping effectiveness, and servo tuning shortcomings.

To simulate external vibrations, a vibrator was attached to the motion stage, inducing controlled vibrations. Moreover, poor servo tuning was investigated by importing incorrect PID parameters into the servo. Fast Fourier Transform (FFT) analysis identified the frequency components of the resulting vibrations. The results showed induced vibrations differ in frequency from those observed at the endpoint. As shown in Figure 8, improper PID settings caused vibrations with a peak frequency of 98 Hz, differing from endpoint vibrations.

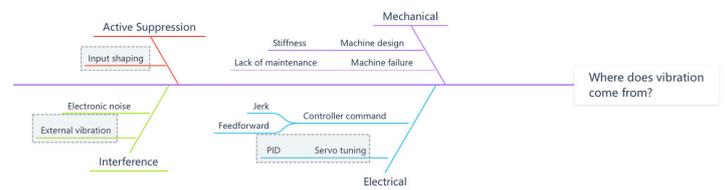

Figure 7: Cause and effect diagram

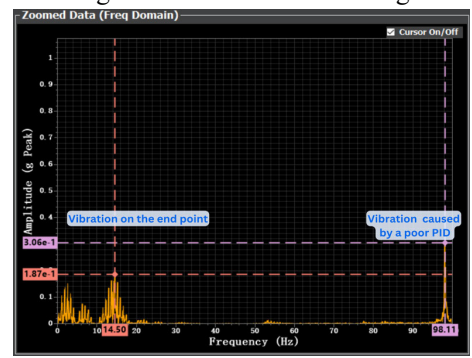

Figure 8: FFT diagram of different vibration

## 4.4 Improve Stage

The Improve stage of DMAIC focuses on implementing solutions to address root causes identified during the Analyze phase. To address vibration issues affecting stage performance and UPH, we used a Design of Experiments (DOE) approach to optimize factors influencing vibration as shown in Table 1 and 2. Using the FrF2 library in R Studio, we analyzed and fit a linear regression model with interaction effects, shown in Figure 9. The p-values for Active Suppression, Servo PID, and their interaction term are all less than 0.05, indicating statistical significance, while External Vibration is not. Active Suppression has a stronger influence on vibration than Servo PID.

Table 1: Factor and level table in DOE.

| Factor | LEVEL | |
| --- | --- | --- |
| | -1 | 1 |
| Active Suppression | NO | YES |
| External Vibration | NO | YES |
| Servo Tuning PID | GOOD | BAD |

Table 2: Experiment conduction in DOE.

| Trail | Active Suppression | Servo Tuning PID | External Vibration | RMS |
| --- | --- | --- | --- | --- |
| 1 | -1 | -1 | -1 | 0.5238 |
| 2 | -1 | -1 | +1 | 0.4977 |
| 3 | -1 | +1 | -1 | 0.579 |
| 4 | -1 | +1 | +1 | 0.6228 |
| 5 | +1 | -1 | -1 | 0.2969 |
| 6 | +1 | -1 | +1 | 0.2997 |
| 7 | +1 | +1 | -1 | 0.486 |
| 8 | +1 | +1 | +1 | 0.4418 |

Figure 10 supports this with main and interaction effects plots. Parallel lines in the interaction plot suggest the effects of Active Suppression and Servo PID are independent.

```
Coefficients:
                  Estimate Std. Error t value Pr(>|t|)    
(Intercept)       0.468463   0.004871  96.177 7.20e-15 ***
ActiveS1         -0.087363   0.004871 -17.936 2.37e-08 ***
ExtVib1          -0.002963   0.004871  -0.608  0.55808    
ServoPID1         0.063937   0.004871  13.127 3.57e-07 ***
ActiveS1:ExtVib1 -0.007387   0.004871  -1.517  0.16366    
ActiveS1:ServoPID1 0.018863  0.004871   3.873  0.00377 **
ExtVib1:ServoPID1 0.002863   0.004871   0.588  0.57119    
```

Figure 9: P value of the factors

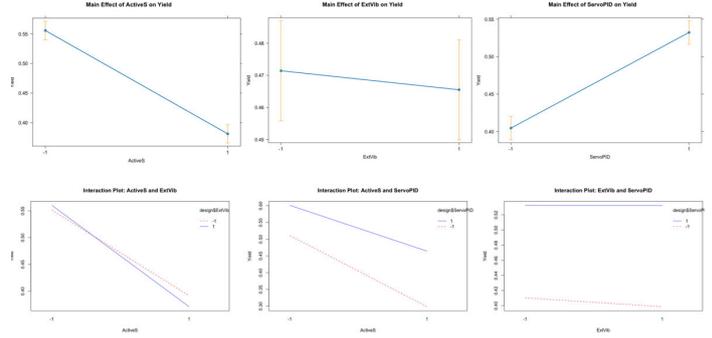

Figure 10: Effect plots of the factors

The limited effect of external vibration is due to the external vibrator's attachment to the stages, preventing direct endpoint influence. In production, passive vibration suppression further minimizes its impact. Our test demonstrated that active vibration suppression through input shaping can reduce endpoint vibration by 25-40%. Proper Servo PID tuning is also crucial during stage assembly and fine-tuning after installation.

## 4.5 Control Stage

To ensure sustained vibration mitigation, X Corporation is introducing VIBGUARD, a comprehensive solution that includes a Soft-motion controller, accelerometers, an EtherCAT sensor module, servo drives, and motion stages. VIBGUARD performs the following critical functions:

- Endpoint and Stage Vibration Analysis: Analyzes vibrations on both the endpoint and the stage, enabling proactive mitigation.
- Active Vibration Suppression: Actively suppresses vibrations to optimize machine performance.
- Real-time Vibration Monitoring: Continuously monitors vibration levels for immediate anomaly detection.
- Alarm and Warning Systems: Triggers alarms or warnings in response to sudden or excessive vibrations, prompting swift corrective actions by FAB operators.

This integrated approach caters to the competitive market landscape by providing value-added solutions. Additionally, X Corporation is incorporating vibration analysis and mitigation into its internal process map to simulate practical motion settings, as shown in Figure 11. Following VIBGUARD's implementation, UPH throughput increased by 15.3%, from 8,500 to 9,800 units, and downtime decreased by 68.2%, with monthly occurrences dropping from 2.2 to 0.7. During implementation, challenges such as system integration and calibration were encountered. These were addressed through rigorous testing and collaboration with technical experts to ensure seamless operation.

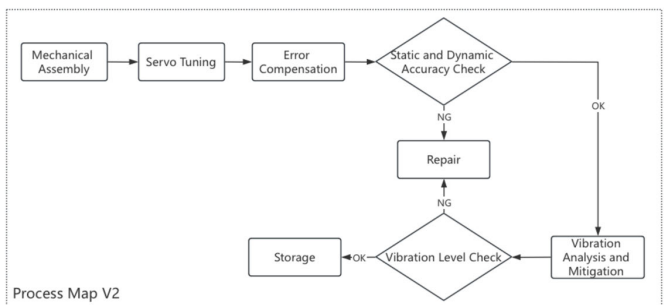

Figure 11: Process map V2

## 5. CONCLUSION

This study demonstrated the application of the DMAIC methodology to identify and mitigate vibration issues in semiconductor manufacturing. We identified primary vibration causes, such as external sources, input shaping effectiveness, and servo tuning quality. Implementing active vibration suppression and proper servo PID tuning significantly improved performance. The VIBGUARD solution enhanced productivity by integrating real-time vibration monitoring and suppression, boosting UPH by 15.3% and reducing downtime by 68.2%.

However, this study has limitations, including the potential variability of vibration sources and the generalizability of the results to different manufacturing settings. Future research should explore the long-term impacts of these interventions and investigate other advanced vibration control techniques to further optimize semiconductor manufacturing processes. This underscores the importance of a proactive strategy in enhancing value and throughput in demanding environments.

**ACKNOWLEDGMENTS**

This work is funded by National Natural Science Foundation of China Grant No. 72371271, the Guangzhou Industrial Information and Intelligent Key Laboratory Project (No. 2024A03J0628), the Nansha Key Area Science and Technology Project (No. 2023ZD003), and Project No. 2021JC02X191. This work is also supported by grants from the Shenzhen University-Enterprise Joint Research and Development Project: The Development of High-speed and High-precision Analysis Test Platform Based on Software Motion Control for the Semiconductor Industry (Grant Nos. 2021030555401610).